\documentclass[twocolumn,10pt]{svjour3}
\usepackage{svjour3fix}
\let\email=\relax
\usepackage[T1]{fontenc}
\usepackage[charter]{mathdesign}

\usepackage[utf8]{inputenc}
\usepackage[breaklinks,colorlinks=true,ps2pdf=true]{hyperref}
\usepackage[dvips]{graphicx}
\usepackage{psfrag}
\usepackage[english]{babel}
\usepackage{xspace}
\usepackage{breakurl}
\usepackage{bodyversion}

\newcommand{\term}[1]{\emph{#1}}		
\newcommand{\etal}{\textit{et al.}~}	
\newcommand{\ie}{{i.e.},\ }		

\def\figscale{0.5}
\newcommand{\pds}[1]{\textsc{#1}}
\newcommand{\pdp}[1]{\textit{#1}}
\newcommand{\ppr}[1]{\textit{#1}}
\newcommand{\pet}[1]{\textrm{`#1'}}

\advance\oddsidemargin by 9pt
\advance\evensidemargin by 9pt

\date{\today}

\title{Privacy Design Strategies}

\author{Jaap-Henk Hoepman}
\institute{
Jaap-Henk Hoepman \at
TNO\\
  P.O. Box 1416, 9701 BK \ Groningen, The Netherlands\\
  \email{jaap-henk.hoepman@tno.nl}\\
  and\\
  Institute for Computing and Information Sciences (ICIS)\\
  Radboud University Nijmegen\\
  P.O. Box 9010, 6500 GL \ Nijmegen, the Netherlands\\ 
  \email{jhh@cs.ru.nl}.
}

\begin{document}

\maketitle

\hyphenation{de-mon-stra-te}
\hyphenation{con-trol-ler}
\hyphenation{arch-es}
\hyphenation{whe-ther}
\hyphenation{know-led-ge}
\hyphenation{pre-ser-ving}
\hyphenation{Swee-ney}
\hyphenation{sour-ces}
\hyphenation{data-bases}
\hyphenation{favour-ab-le}
\hyphenation{now-a-days}
\hyphenation{trans-pa-ren-cy}
\hyphenation{breach-es}
\hyphenation{com-po-nents}
\hyphenation{dis-tin-guish-ed}
\hyphenation{there-by}
\hyphenation{stra-te-gies}
\hyphenation{pu-re-ly}
\hyphenation{a-na-ly-sis}
\hyphenation{draft-ed}
\hyphenation{hard-ly}
\hyphenation{a-chie-ves}

\begin{abstract}
In this paper we define the notion of a privacy design strategy. These strategies help IT architects to support privacy by design early in the software development life cycle, during concept development and analysis. Using current data protection legislation as point of departure we derive the following eight privacy design strategies: \pds{minimise}, \pds{hide}, \pds{separate}, \pds{aggregate}, \pds{inform}, \pds{control}, \pds{enforce}, and \pds{demonstrate}. The strategies also provide a useful classification of privacy design patterns and the underlying privacy enhancing technologies. We therefore believe that these privacy design strategies are not only useful when designing privacy friendly systems, but also helpful when evaluating the privacy impact of existing IT systems. 
\end{abstract}

\section{Introduction}

The goal of privacy\footnote{%
  In this paper we focus on data protection, and treat privacy and
  data-protection as synonyms.
} 
by design is to take privacy requirements into account
throughout the system development process, from the conception of a new IT system up to its realisation~\cite{cavoukian2011pbd}. The underlying motivation for this approach is that by taking privacy serious from the start the final system will be more privacy friendly. Privacy by design is becoming more important. For example, the proposal for a new European data protection regulation~\cite{COM(2012)11} explicitly requires data protection by design and by default. It is therefore crucial to support developers in satisfying these requirements with practical tools and guidelines.

In the context of developing IT systems, privacy by design implies that privacy protection is a system requirement that must be treated like any other functional requirement. As a result, also privacy protection will have an impact on the design and implementation of the system. To support privacy by design, we therefore need guiding principles to support the inclusion of privacy requirements throughout the system development life cycle

But there is a considerable gap to be bridged here: so far there is little experience in applying privacy by design in engineering~\cite{gurses2011engineering-privacy}. 

As explained in Section~\ref{sec-development}, an important methodology during the design phase is the application of so called software design patterns. These design patterns refine the system architecture to achieve certain functional requirements within a given set of constraints. 
During software development the availability of practical methods to protect privacy is high during actual implementation, but low when starting the project. Numerous privacy enhancing technologies (PETs) exists that can be applied more or less 'of the shelf'. Before that implementation stage, privacy design patterns can be used during system design. Significantly less design patterns exist compared to PETs, however. And at the start of the project, during the concept development and analysis phases, the developer stands basically empty handed. 

This paper aims to contribute to closing this gap. Design patterns do not necessarily play a role in the earlier, concept development and analysis, phases of the software development cycle. The main reason is that such design patterns are already quite detailed in nature, and more geared towards solving an implementation problem. To guide the development team in the earlier stages, we define the notion of a privacy design strategy. Because these strategies describe fundamental approaches to protecting privacy, they enable the IT developer to make well founded choices during the concept development and analysis phase. These choices have a huge impact on the overall privacy protection properties of the final system.

The privacy design strategies developed in this paper are derived from existing privacy principles and data protection laws. These are described in section~\ref{sec-legal}. We focus on the principles and laws, on which the design of an IT system has a potential impact. By taking an abstract information storage model of an IT system as a point of departure, 
these legal principles are translated to a context more relevant for the IT developer in  section~\ref{sec-determine}. This leads us to define the following privacy design strategies: \pds{minimise}, \pds{hide}, \pds{separate}, \pds{aggregate}, \pds{inform}, \pds{control},  \pds{enforce} and \pds{demonstrate}. They are described in detail in section~\ref{sec-pds}.
We validate our approach in section~\ref{sec-validate} by verifying that the strategies we derived indeed cover the legal principles on which they are based. We also show that the strategies apply to an information flow type of system. 

We believe these strategies help to support privacy by design throughout the full software development life cycle, even before the design phase. It makes explicit which high level decisions can be made to protect privacy, when the first concepts for a new information system are drafted. 
The strategies also provide a useful classification of privacy design patterns and the underlying privacy enhancing technologies. 
We therefore believe that these privacy design strategies are not only useful when designing privacy friendly systems, but that they also provide a starting point for evaluating the privacy impact of existing information systems.

Our approach builds on the framework by Spiekermann and Cranor~\cite{DBLP:journals/tse/SpiekermannC09} that distinguishes four stages of privacy-friendliness (ranging from fully identified to completely anonymous). In their framework, the two highest stages are achieved through a privacy-by-architecture approach. The two lowest stages require
a privacy-by-policy approach. They see these two approached as essentially mutually exclusive: ``In contrast, if companies do not opt for a
privacy-by-architecture approach, then a privacy-by-policy approach must be taken where notice and choice will be essential mechanisms for ensuring adequate privacy protection''~\cite{DBLP:journals/tse/SpiekermannC09}. In other words, a system that is engineered as privacy-by-architecture does not process privacy sensitive data and therefore does not need privacy-by-policy.

Our view is less binary: a system architecture will hardly ever guarantee full privacy, and a privacy policy alone does not give sufficient privacy guarantees either. We aim to provide system designers concrete strategies to actually engineer privacy, from both perspectives. Some of these strategies cover the privacy-by-architecture approach. Others cover the privacy-by-policy approach. These strategies are not mutually exclusive, however. Any subset of the strategies can be applied in parallel when designing a system in a privacy friendly manner.

\section{Software development}
\label{sec-development}

Software architecture encompasses the set of significant decisions about the organisation of a software system\footnote{%
  Based on an original definiton by Mary Shaw, expanded in 1995 by Grady
  Booch, Kurt Bittner, Philippe Kruchten and Rich Reitman as reported
  in~\cite{kruchten2004ontology}.
}, 
including
\begin{itemize}
\item the selection of the structural elements and their interfaces by which a
   system is composed,
\item the behaviour as specified in collaborations among those elements,
\item the composition of these structural and behavioural elements into larger
    subsystem, and
\item the architectural style that guides this organisation.
\end{itemize}

\begin{figure}[t]
\begin{center}
\includegraphics[scale=\figscale]{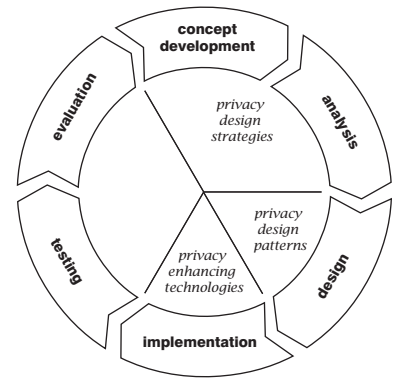}
\end{center}
\caption{The software development cycle and the relationship with strategies, patterns and technologies.}
\label{fig-cycle}
\end{figure}

Typically, the development of a software systems proceeds in six phases: concept development, analysis, design, implementation, testing and evaluation. In fact, these phases are often considered a cycle, where after evaluation a new iteration starts by updating the concept as appropriate.
In this paper we distinguish design strategies (defined in this paper), design patterns and concrete (privacy enhancing) technologies as tools to support the decisions to be made in each of these phases. The design strategies support the concept development and analysis phases, the design patterns are applicable during the design phase, and the privacy enhancing technologies are useful during the implementation phase. This is depicted in figure~\ref{fig-cycle}.

\subsection{Design patterns}

The concept of a \term{design pattern} is a useful vehicle for making design  decisions about the organisation of a software system. A design pattern
\begin{quote}
``provides a scheme for refining the subsystems or components of a software system, or the relationships between them. It describes a commonly recurring structure of communicating components that solves a general design problem within a particular context.''~\cite{buschmann1996patterns} 
\end{quote}
Typically, the description~\cite{gamma1994design-patterns} of a design pattern contains at least its name, purpose, context (the situations in which it applies), implementation (its structure, components and their relationships), and the consequences (its results, side effects and trade offs when applied). Many design patterns exist, at varying levels of abstraction. 

People familiar with classical architecture may find it useful to compare a design pattern with common structural approaches in the construction of buildings. For example, an arch is a typical structure that spans a certain space and supports the structure and mass above it. Arches are used for this purpose in many different types of buildings and constructions, ranging from aqueducts to cathedrals. This is why an arch can be considered a design pattern. 

A classical software design pattern is the \pdp{Model-View-Controller}\footnote{%
  Originally formulated in the late 1970s by Trygve Reenskaug at Xerox PARC,
   as part of the Smalltalk system.
}, 
that separates the representation of the data (the model) from the way it is represented towards the user (the view) and how the user can modify that data (using the controller). A much simpler design pattern is the \pdp{Iterator} pattern, that ``provides a way to access the elements of an aggregate object sequentially without exposing its underlying representation''~\cite{gamma1994design-patterns}. By using this pattern, the actual implementation of the list to process has become irrelevant and can be changed without changing higher level code.

Few privacy design patterns have been explicitly described as such to date. We are aware of the work of Hafiz~\cite{hafiz2006pdp,hafiz2011patternlanguage}, Pearson~\cite{DBLP:conf/trustbus/PearsonS10,pearson2010decision-support},
van Rest~\etal\cite{rest2012designing-privacy}, and a recent initiative of the UC Berkeley School of Information\footnote{%
  \url{http://privacypatterns.org/}
}.
Many more privacy design patterns exist though, although they have never been described as such. Sweeney's \pdp{$k$-anonymity} concept~\cite{DBLP:journals/ijufks/Sweene02} is a classical example of an idea that implicitly defines a privacy design pattern. Also the concept of a \pdp{zero knowledge proof}~\cite{DBLP:journals/siamcomp/GoldwasserMR89} can be viewed as a design pattern. The same can be said of \pdp{mix networks}, based on the concept op onion routing~\cite{DBLP:journals/cacm/Chaum81}. 
Moreover, many privacy enhancing technologies implicitly define a corresponding privacy design pattern. Good examples are \pdp{attribute based credentials} based on \pet{Idemix}~\cite{DBLP:conf/eurocrypt/CamenischL01} and \pet{u-prove}~\cite{brands2000pki}, and studied in for example the IRMA\footnote{%
  \url{http://www.irmacard.org}
}
and the ABC4TRUST project\footnote{%
  \url{http://www.abc4trust.eu}
}.

\subsection{Design strategies}

Because certain design patterns have a higher level of abstraction than others, some authors also distinguish \term{architecture patterns}, that
\begin{quote}
``express a fundamental structural organisation or schema for software systems. They provide a set of predefined subsystems, specify their responsibilities, and include rules and guidelines for organising the relationships between them.''\footnote{%
See \url{http://best-practice-software-engineering.ifs.tuwien.ac.at/patterns.html},
and The Open Group Architecture Framework (TOGAF)\\ \url{http://pubs.opengroup.org/architecture/togaf8-doc/arch/chap28.html}
}
\end{quote}
The \pdp{Model-View-Controller} pattern cited above is often considered such an architecture pattern. The distinction between an architecture pattern and a design pattern is not always easily made, however. Moreover, there are even more general principles that guide the system architecture without imposing a specific structural organisation or schema for the system. 

We choose, therefore, to express such higher level abstractions in terms of \term{design strategies}. We define this as follows.
\begin{quote}
A design strategy describes a fundamental approach to achieve a certain design goal. It has certain properties that allow it to be distinguished from other (fundamental) approaches that achieve the same goal.
\end{quote}
Whether something classifies as a strategy very much depends on the universe of discourse, and in particular on the exact goal the strategy aims to achieve. For example, if the goal is to cross a river, possible strategies are \pds{to build a bridge}, \pds{use the ferry}, or \pds{learn to fly}. However, if the goal is to build a bridge, then the discourse changes. The construction of a bridge may be classified depending on how the forces of tension, compression, bending, torsion and shear are distributed through its structure. A very strategic decision is to decide whether to use a form of \pds{suspension} (where the deck is suspended from below a main cable) instead of using a more classical form of \pds{support} (for example, using arches).

A \term{privacy design strategy} is a design strategy that achieves (some level of) privacy protection as its goal. 

Design strategies do not necessarily impose a specific structure on the system although they certainly limit the possible structural realisations of it. Therefore, they are also applicable during the concept development and analysis phase of the development cycle\footnote{%
  We note that the notion of a privacy design strategy should not be confused
  with the foundational principles of Cavoukian~\cite{cavoukian2011pbd}
  or the concept of a privacy principle from the ISO 29100 Privacy
  framework~\cite{ISO29100}.
}.

\subsection{Privacy enhancing technologies}

\term{Privacy Enhancing Technologies} (PETs) are better known, and much more studied. Borking and Blarkom~\etal\cite{borking1996identityprotector,blarkom2003pet} define them as follows.
\begin{quote}
``Privacy-Enhancing Technologies is a system of ICT measures protecting informational privacy by eliminating or minimising personal data thereby preventing unnecessary or unwanted processing of personal data, without the loss of the functionality of the information system.''
\end{quote}
This definition was later adopted almost literally by the European Commission~\cite{COM(2007)228}. It is slightly biased towards the data-minimisation principle, and it suggests that privacy enhancing technologies are slightly more high-level than those that are typically studied. 

In principle, PETs are used to implement a certain privacy design pattern with concrete technology. For example, both \pet{Idemix}~\cite{DBLP:conf/eurocrypt/CamenischL01} and \pet{u-prove}~\cite{brands2000pki} are privacy enhancing technologies implementing the (implicit) design pattern~\pdp{anonymous credentials}.
There are many more examples of privacy enhancing technologies, like \pet{cut-and-choose} techniques~\cite{DBLP:conf/crypto/ChaumFN88}, 
\pet{onion routing}\footnote{%
  Made popular through the TOR project \url{http://www.torproject.org/}.
}~\cite{DBLP:journals/cacm/Chaum81} to name but a few. Other good sources for privacy enhancing technologies are the survey of Ian Goldberg~\cite{DBLP:conf/pet/Goldberg02}, the Stanford PET wiki\footnote{%
  \url{http://cyberlaw.stanford.edu/wiki/index.php/PET}
}, and the proceedings of the annual Privacy Enhancing Technologies Symposium\footnote{%
  \url{http://petsymposium.org/}
}.

\subsection{Discussion}

The distinction between a design strategy, a design pattern and a privacy enhancing technology are in some way similar to the classification of strategic, tactical and operational levels of decision making commonly found in the management literature. Decisions at the strategic level are aimed to fulfil the company's mission (in this case 'achieve privacy'). They are not specific to particular organisational units, and typically express 'what' to achieve, without paying much attention to or giving details on 'how' to achieve this. This corresponds to our notion of a privacy design strategy.
Decisions at the tactical level make the strategic decisions more concrete, and translate them to goals and plans for a particular organisational unit. This corresponds to our notion of a design pattern that helps solving a generic problem within a particular context. Operational decisions implement the tactical plans by assigning resources, scheduling production, etc. This
correspond to our notion of (a combination of) privacy enhancing technologies that implement a privacy design pattern.

An earlier draft of this paper confused anonymous credentials for a privacy enhancing technology. This is wrong. In fact \pdp{attribute based credentials}\footnote{%
  Attribute based credentials is a better term than anonymous credentials
  because in many cases the credential may contain non-anonymous information. 
}
are a perfect example of a privacy design pattern\footnote{%
  In this paper we will occasionally refer to a design pattern as if it is
  already properly defined as such. This is often not the case (including the
  case of the \pdp{attribute based credential} at hand here). For now we 
  will just appeal to the intuitive understanding of the main structure of the
  pattern, and defer a full description of such a pattern to further research.
  In a way, the secondary purpose of this paper is to identity such new privacy
  design patterns, and to merely record their existence for now.
}. 
This pattern describes a general structure separating users, issuers, and verifiers, where the link between issuing a credential and proving possession of it is broken to prevent tracking users. This is a telling testimony to the fact that the distinction between a privacy design pattern and a privacy enhancing technology is not so easy to make in practice (if only because historically many design patterns are only implicitly defined by the corresponding privacy enhancing technology).

Let us try to make the distinction clearer by reconsidering the example of the construction of a bridge introduced above. In terms of this example, a design strategy is the use of \pds{support} (instead of \pds{suspension}). After choosing the \pds{support} strategy, there are several design options one can apply. One of these options is the use of arches to create the structural support required. \pdp{Arches} then is a design pattern --- in fact a design pattern that occurs in many different constructions beyond bridges. A concrete technology to build (\ie implement) an arch bridge is to use bricks for the actual construction of a so-called \pet{round arch}.

Design patterns may overlap, and may vary in the level of detail they provide. Similar to the difference in abstraction between the \pdp{Model-View-Controller} and the \pdp{Iterator} pattern, the \pdp{attribute based credentials} pattern is much more concrete than a more generic \pdp{use pseudonyms} pattern, and in fact may partially overlap with that more generic pattern. Finally, we note that a privacy design pattern may sometimes implement several privacy design strategies.


\section{The foundations of data protection}
\label{sec-legal}

We aim to derive privacy design strategies from existing data protection laws and privacy frameworks. We therefore briefly summarise those here.

In the European Union, the legal right to privacy is based on Article 8 of the European Convention of Human Rights of 1950. In the context of data protection, this right has been made explicit in the 1995 data protection directive~\cite{ec-95-46}, which is based on the privacy guidelines of the Organisation of Economic Co-Operation and Development (OECD) from 1980~\cite{oecd1980guidelines}. 

\subsection{The OECD guidelines}

The OECD guidelines, of which the US fair information practices (FIPs)~\cite{ftc2000fips} --- 
\emph{notice}, \emph{choice}, \emph{access} and \emph{security} ---
are a subset, define the following principles.
\begin{itemize}
\item The collection of personal data is lawful, limited, and happens with
the knowledge or consent of the data subject (Collection Limitation).
\item Personal data should be relevant to the purposes for which they are to
be used, and be accurate, complete and kept up-to-date (Data Quality).
\item The purposes of the collection must be specified upfront  (Purpose Specification), and the use of the data after collection is limited to that purpose (Use Limitation).
\item Personal data must be adequately protected (Security Safeguards).
\item The nature and extent of the data processing and the controller responsible must be readily available (Openness).
\item Individuals have the right to view, erase, rectify, complete or amend personal data stored that relates to him (Individual Participation).
\item A data controller must be accountable for complying with these principles (Accountability).
\end{itemize}

\subsection{Data protection in Europe}

The OECD principles correspond roughly to the main provisions in the European data protection Directive of 1995~\cite{ec-95-46}. For example, Article 6 states that personal data must be
processed fairly and lawfully, must be collected for a specified purpose, and must not be further processed in a way incompatible with those purposes. Moreover the data must be adequate, relevant, and not excessive. It must be accurate and up to date, and kept no longer than necessary. These provisions express a need for purpose limitation, data minimisation, and data quality.

Other articles of the Directive deal with transparency and user choice. For example, article 7 requires unambiguous consent from the data subject, while article 10 and 11 require data controllers to inform data subjects about the processing of personal data. Article 12 gives data subjects the right to review and correct the personal data that is being processed about them. Finally, security as a means to protect privacy is addressed by article 17, that mandates adequate security of processing.

We note that the European data protection directive covers many more aspects, that are however less relevant for the discussion in this paper. These aspects are not concerned with how a system is designed or how it operates, but are more concerned with the organisational embedding of the system for example.

The directive is currently under review and a proposal for a regulation to replace it has recently been published~\cite{COM(2012)11}. This regulation is still in flux and under heavy debate, but it contains the following rights and obligations that are relevant for our discussion in this paper.
\begin{itemize}
\item A controller must implement data protection by design and by default (article 23).
\item A controller must be able to demonstrate compliance with the regulation (article 5, and also article 22).
\item Data subjects have the right to be forgotten and to erasure (article 17).
\item Data subjects have the right to data portability, allowing them `to obtain from the controller a copy of data undergoing processing in an electronic and structured format which is commonly used and allows for further use by the data subject' (article 18).
\item A data controller has the duty to issue a notification whenever a personal data breach occurs (article 31 and 32).
\end{itemize}

\subsection{The ISO 29100 Privacy Framework perspective}

In response to the growing importance of privacy by design, 
the International Organisation for Standardisation (ISO) issued the ISO 29100 Privacy framework~\cite{ISO29100}. This framework collects organisational, technical and procedural aspects of privacy protection, with the intention to enhance existing security standards. The ISO 29100 principles lie somewhere between purely legal requirements, and the more technically oriented design strategies that we aim to develop here. The standard
suggests the following eleven privacy principles.
\begin{itemize}
\item \ppr{Consent and choice}: inform data subjects, present the available choices and obtain consent.
\item \ppr{Purpose legitimacy and specification}: ensure compliance with data protection legislation and inform data subjects.
\item \ppr{Collection limitation}: limit data collection to what is needed for the purpose.
\item \ppr{Data minimisation}: minimise the amount of personal data collected, minimise the number of actors that have access, offer as default non-privacy invasive options, and delete data once it has become no longer necessary.
\item \ppr{Use, retention and disclosure limitation}: limit the use, retention and disclosure of personal data to what is needed for the purpose.
\item \ppr{Accuracy and quality}: ensure the data is accurate, up-to-date, adequate and relevant, verify this, and periodically check this.
\item \ppr{Openness, transparency and notice}: inform data subjects about the data controller policies, give proper notices that personal data is being processed, and provide information on how to access and review personal data.
\item \ppr{Individual participation and access}: give data subjects a real possibility to access and review their personal data.
\item \ppr{Accountability}: document policies, procedures and practices, assign the duty to implement privacy policies to specified individuals in the organisation, provide suitable training, inform about privacy breaches, give access to effective sanctions and procedures for compensations in case of privacy breaches.
\item \ppr{Information security}: provide a proper level of security, and implement the right controls, based on an appropriate risk assessment.
\item \ppr{Privacy compliance} verify and demonstrate that the IT systems meets legal requirements, and have appropriate internal controls and supervision mechanisms.
\end{itemize}
We will not discuss the merits of this subdivision in any depth, although we do note that there is quite a bit of overlap between 
\ppr{Consent and choice}, \ppr{Purpose specification} and \ppr{Openness, transparency and notice}. Similarly, \ppr{Collection limitation}, \ppr{Data minimisation} and \ppr{Use, retention and disclosure limitation} all more or less describe the same need for data minimisation. As a result, the principles are only partially useful from the perspective of analysis and design.

\subsection{Summary of requirements}
\label{ssec-sumreq}

Not every legal requirement can be met by designing an IT system in a specific way. Legitimacy of processing is a good example. If the processing is illegitimate, then it will be illegitimate irrespective of the design of the system. We therefore focus our effort on studying aspects on which the design of an IT system has a potential impact, summarised in the list below.
\begin{itemize}
\item Purpose limitation (comprising both specification of the purpose and limiting the use to that stated purpose).
\item Data minimisation.
\item Data quality.
\item Transparency (Openness in OECD terms).
\item Data subject rights (in terms of consent, and the right to view, erase, and rectify personal data).
\item The right to be forgotten.
\item Adequate protection (Security Safeguards in OECD terms).
\item Data portability
\item Data breach notifications.
\item Accountability and (provable) compliance.
\end{itemize}
These principles must be covered by the privacy design strategies that we will derive next. Whether this is indeed the case is analysed in section~\ref{sec-validate}.

\section{Deriving privacy design strategies}
\label{sec-determine}

A natural starting point to derive privacy preserving strategies is to look at when and how privacy is violated, and then consider how these violations can be prevented. Solove's taxonomy~\cite{solove2006taxonomy}, for example, identifies four basic groups of activities that affect privacy: information collection, information processing, information dissemination and invasions.
This is in fact similar to the distinction made between data transfer, storage and processing by Spiekermann and Cranor~\cite{DBLP:journals/tse/SpiekermannC09}. He then discusses in detail the different ways in which certain specific activities (like surveillance, aggregation, disclosure, and intrusion) affect privacy. The specific activities identified by Solove are too fine grained. Although they may in fact be interesting to distinguish from a legal perspective, many of them involve basically the same methods at a technical level. His general subdivision however inspired us to look at IT systems at a higher level of abstraction to determine where and how privacy violations can be prevented.

In doing so, we can view an IT system either as an information storage system (\ie database system) or an information flow system. Many of today's systems, like classical business or government administration systems, are database systems. The same holds for social networks. An alternative view of an IT system is that of an information flow system. This view makes sense if the sheer volume of data (for example those created by sensor networks or the Internet of Things) becomes too large to store. Interestingly, as we shall see later, both views on IT systems are subject to the same eight privacy design strategies. We note that in practice a system is often a hybrid, with some information flow components and some information storage components.

Let us first consider the information storage system view. Current data protection legislation~\cite{ec-95-46} is pretty much written with such a database model in mind.
In a database, information about individuals is stored in one or more tables. Each table
stores a fixed set of attributes for an individual. The columns in the table represent this fixed set of attributes. A row is added for each new individual about whom a record
needs to be stored. Sometimes, data is not stored at the level of individual persons, but is instead aggregated based on certain relevant group properties (like postal code). 

Within the legal framework described in section~\ref{sec-legal}, the collection of personal data should be proportional to the purpose for which it is collected, and this purpose should not be achievable through other, less invasive means. In practice, this means that data collection should be \emph{minimised}. This can be achieved by not storing individual rows in a database table for each and every individual. Also the collection of attributes stored should correspond to the purpose. Data collected for one purpose should be stored \emph{separately} from data stored for another purpose. Linking of these database tables should not be easy. When data about specific individuals is not necessary for the purpose, only \emph{aggregate} data should be stored. Personal data should be properly protected, and \emph{hidden} from other parties. A data subject should be \emph{informed} about the fact that data about her is being processed, and she should be able to request modifications and corrections where appropriate. In fact the underlying principle of information self-determination dictates the she should be in \emph{control}. Finally, the collection and processing of personal data should be done in accordance to a privacy policy, that should be actively \emph{enforced}. The current proposal for the revision of the European privacy directive (into a regulation) also stresses the fact that data controllers should be able to \emph{demonstrate compliance} with data protection legislation. The data controller has the burden of proof with respect to compliance, and must, for example run and document a privacy impact assessment (PIA).

\begin{figure}[t]
\begin{center}
\includegraphics[scale=\figscale]{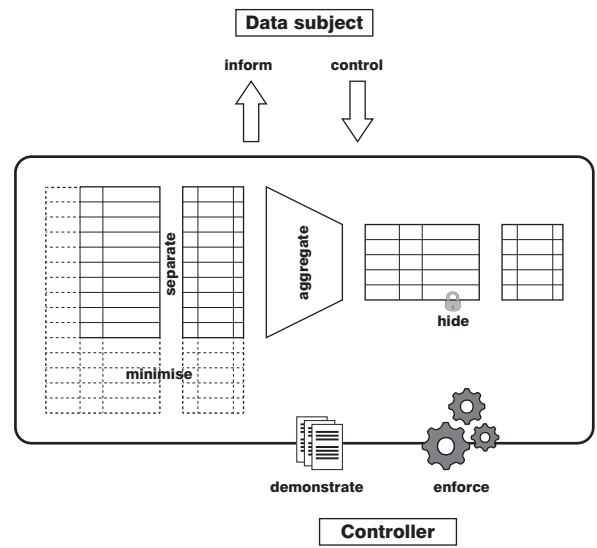}
\end{center}
\caption{The database metaphor of the eight privacy design strategies.} 
\label{fig-db-metaphor}
\end{figure}

Given this analysis form the legal point of view, we distinguish the following eight privacy design strategies: \pds{minimise}, \pds{separate}, \pds{aggregate}, \pds{hide}, \pds{inform}, \pds{control}, \pds{enforce} and \pds{demonstrate}. A graphical representation of these strategies, when applied to a database system, is given in Figure~\ref{fig-db-metaphor}. 

\section{The eight privacy design strategies}
\label{sec-pds}

We will now proceed to describe these eight strategies in detail. 
We have grouped the strategies into two classes: data-oriented strategies and process-oriented strategies. The first class roughly corresponds to the 
privacy-by-archi\-tecture approach identified by Spiekermann and Cranor~\cite{DBLP:journals/tse/SpiekermannC09}, whereas the process-oriented strategies more-or-less cover their privacy-by-policy approach.
For each strategy we define it, explain the definition and outline its scope by giving examples. We also present some design patterns that fit the particular strategy. We would like to stress that the strategies are not mutually exclusive. Any subset of the strategies can be applied in parallel when designing a system in a privacy friendly manner.

\subsection{Data oriented strategies}

\subsubsection{Strategy \#1: \pds{minimise}}

The most basic privacy design strategy is \pds{minimise}, which states that 
\begin{quote}
The amount of personal data that is processed\footnote{%
  For brevity, and in line with Article 2 of the European  
  directive~\cite{ec-95-46}, we use data processing to include 
  the collection, storage and dissemination of that data as well.
} 
should be restricted to the minimal amount possible.
\end{quote}
This strategy is extensively discussed by Gürses~\etal\cite{gurses2011engineering-privacy}. By ensuring that no, or no unnecessary, data is collected, the possible privacy impact of a system is limited. Applying the \pds{minimise} strategy means one has to answer whether the processing of personal data is proportional (with respect to the purpose) and whether no other, less invasive, means exist to achieve the same purpose. The decision to collect personal data can be made at design time and at run time, and can take various forms.
For example, one can decide not to collect any information about a particular data subject at all. Alternatively, one can decide to collect only a limited set of attributes.

\paragraph{Design patterns}

Common design patterns that implements this strategy are \pdp{select before you collect}~\cite{jacobs2005select} , \pdp{anonymisation} and \pdp{use pseudonyms}~\cite{pfitzmann2010privacy-terminology}.

\subsubsection{Strategy \#2: \pds{hide}} 


The second design strategy, \pds{hide}, states that 
\begin{quote}
Any personal data, and their interrelationships, should be hidden from plain view. 
\end{quote}
The rationale behind this strategy is that by hiding personal data from plain view, it cannot easily be abused. The strategy does not directly say from whom the data should be hidden. And this depends on the specific context in which this strategy is applied. In certain cases, where the strategy is used to hide information that spontaneously emerges from the use of a system (for example communication patterns), the intent is to hide the information from anybody. In other cases, where information is collected, stored or processed legitimately by one party, the intent is to hide the information from any other party. In this case, the strategy corresponds to ensuring confidentiality. 

The \pds{hide} strategy is important, and often overlooked. In the past, many systems have been designed using innocuous identifiers that later turned out to be privacy nightmares. Examples are identifiers on RFID tags, wireless network identifiers, and even IP addresses. The \pds{hide} strategy forces one to rethink the use of such identifiers. In essence, the \pds{hide} strategy aims to achieve unlinkability and unobservability~\cite{pfitzmann2010privacy-terminology}. Unlinkability in this context ensures that two events cannot be related to one another (where events can be understood to include data subjects doing something, as well as data items that occur as the result of an event). 

\paragraph{Design patterns}

The design patterns that belong to the \pds{hide} strategy are a mixed bag. One of them is the use of \pdp{encryption} of data (when stored, or when in transit). Other examples are \pdp{mix networks}~\cite{DBLP:journals/cacm/Chaum81} to hide traffic patterns~\cite{DBLP:journals/cacm/Chaum81}, or techniques to unlink certain related events like \pdp{attribute based credentials}~\cite{DBLP:conf/eurocrypt/CamenischL01}, \pdp{anonymisation} and the use of \pdp{pseudonyms}. Note that the latter two patterns also belong to the \pds{minimise} strategy. In the context of databases, adding noise, perturbing data, and hiding patterns in databases~\cite{ilprints1014}
based on \pdp{Differential privacy}~\cite{DBLP:conf/icalp/Dwork06} are some other examples. Also the TrackMeNot browser plugin\footnote{%
\url{http://cs.nyu.edu/trackmenot/}
}, described in~\cite{howe2008trackmenot}, belongs to this class.
Moreover, \pdp{anonymisation} and \pdp{use pseudonyms} (that also belong to the \pds{Minimise} startegy belong to this strategy as well, as they contribute to achieve unlinkability.

For RFID tags, protocols exist that only reveal the identity of the tag to a small set of trusted readers~\cite{juels2006rfid-secpriv-survey}. More generally it would be good practice in wireless network design to verify the authenticity of the access point before trying to connect to it (and thus making yourself known). This is similar to the common pattern found in nature where cubs hide and only reveal themselves after they recognise their mother (by sound, or scent). 

\subsubsection{Strategy \#3: \pds{separate}}

The third design strategy, \pds{separate}, states that 
\begin{quote}
Personal data should be processed in a distributed fashion, in separate compartments whenever possible.
\end{quote}
By separating the processing or storage of several sources of personal data that belong to the same person, complete profiles of one person cannot be made. Moreover, separation is a good method to achieve purpose limitation. The strategy of separation calls for distributed processing instead of centralised solutions. In particular, data from separate sources should be stored in separate databases, and these databases should not be linked. Data should be processed locally whenever possible, and stored locally if feasible as well. Database tables should be split when possible. Rows in these tables should be hard to link to each other, for example by removing any identifiers, or using table specific pseudonyms.

These days, with an emphasis on centralised web based services this strategy is often disregarded. However, the privacy guarantees offered by peer-to-peer networks are considerable. Decentralised social networks like Diaspora\footnote{%
  \url{http://diasporafoundation.org/}
}
are inherently more privacy friendly than centralised approaches like Facebook and Google+.

\paragraph{Design patterns}

No specific design patterns for this strategy are known.  Therefore, further investigations into design pattern that implement the \pds{separate} strategy are required. Especially those that also satisfy business needs, as those typically lead to a centralised solution.

Note that we do not consider \pdp{Access control} to implement the \pds{separate} strategy. \pds{Separate} is a data-oriented strategy. It is only concerned with separating data itself, and not concerned with separating people with access to the data.

\subsubsection{Strategy \#4: \pds{aggregate}} 

The fourth design pattern, \pds{aggregate}, states that 
\begin{quote}
Personal data should be processed at the highest level of aggregation and with the least possible detail in which it is (still) useful.
\end{quote}
Aggregation of information over groups of attributes or groups of individuals, restricts the amount of detail in the personal data that remains. This data therefore becomes less sensitive. When the information is sufficiently coarse grained, and the size of the group over which it is aggregated is sufficiently large, little information can be attributed to a single person, thus protecting its privacy.

\paragraph{Design patterns}

Examples of design patterns that belong to this strategy are the following.

\pdp{Aggregation over time} is used to provide some level of privacy protection in smart metering and smart grid systems. Instead of recording  energy use in real time, only cumulative values over 15 minute intervals are reported.
\pdp{Dynamic location granularity} is another approach, used in location based services. It adapts the accuracy of the reported location of a user to ensure that a reasonable number of other users are at the same reported location.
\pdp{$k$-anonymity}~\cite{DBLP:journals/ijufks/Sweene02} and  \pdp{$l$-diversity}~\cite{DBLP:journals/tkdd/MachanavajjhalaKGV07}) are also design patterns in this class (although one could argue that these are more a concept that a concrete method to achieve the specified property of being indistinguishable among a set of at least $k$ entities).

\subsection{Process oriented strategies}

\subsubsection{Strategy \#5: \pds{inform}}

The \pds{inform} strategy corresponds to the important notion of transparency:
\begin{quote}
Data subjects should be adequately informed whenever personal data is processed.
\end{quote}
Whenever data subjects use\footnote{%
  Use can be explicit, like in signing up to a certain service, or implicit,
  like entering an area with camera surveillance. This broader understanding
  of 'engaging with' a system is also important in ambient intelligent systems 
  and the Internet of Things~\cite{hoepman2011iot-trust}.
} 
a system, they should be informed about which information is processed, for what purpose, and by which means. This includes information about the ways the information is protected, and being transparent about the security of the system. Providing access to clear design documentation is also a good practice. Data subjects should also be informed about third parties with which information is shared. And data subjects should be informed about their data access rights and how to exercise them.

\paragraph{Design patterns}

A possible design patterns in this category is the \pdp{Platform for Privacy Preferences (P3P)}\footnote{%
  \url{http://www.w3.org/P3P/}
}. 
\pdp{Data breach notifications} are also a design pattern in this category. 
Finally, Graf \etal\cite{graf2010patterncollection} provide an interesting collection of privacy design patterns for informing the user from the Human Computer Interfacing perspective.

\subsubsection{Strategy \#6: \pds{control}}

The control strategy states that 
\begin{quote}
Data subjects should be provided agency over the processing of their personal data.
\end{quote}
The \pds{control} strategy is in fact an important counterpart to the \pds{inform} strategy. Without reasonable means of controlling the use of one's personal data, there is little use in informing a data subject about the fact that personal data is collected. Of course, the converse also holds: without proper information, there is little use in asking consent. Data protection legislation often gives the data subject the right to view, update and even ask the deletion of personal data collected about her. This strategy underlines this fact, and design patterns in this class give users the tools to exert their data protection rights.

\pds{Control} goes beyond the strict implementation of data protection rights, however. It also governs the means by which users can decide whether to use a certain system, and the way they control what kind of information is processed about them. In the context of social networks, for example, the ease with which the user can update his privacy settings through the user interface determines the level of control to a large extent. So user interaction design is an important factor as well. Moreover, by providing users direct control over their own personal data, they are more likely to correct errors. As a result the quality of personal data that is processed may increase.

\paragraph{Design patterns}

We are not aware of specific design patterns that fit this strategy. 
Perhaps experience from user interaction design may be applied to create design patterns for obtaining consent.

\subsubsection{Strategy \#7: \pds{enforce}}

The seventh strategy, \pds{enforce}, states: 
\begin{quote}
A privacy policy compatible with legal requirements should be in place and should be enforced.
\end{quote}
The \pds{enforce} strategy ensures that a privacy policy is in place. This is an important step in ensuring that a system respects privacy during its operation. Of course, the actual level of privacy protection depends on the actual policy. At the very least it should be compatible with legal requirements. As a result, purpose limitation is covered by this strategy as well.
More importantly thought, the policy should be enforced. This implies, at the very least, that proper technical protection mechanisms are in place that prevent violations of the privacy policy. Moreover, appropriate governance structures to enforce that policy must also be established.

\paragraph{Design patterns}

\pdp{Access control} is an example of a design patterns that implement this strategy. Another example is privacy rights management: a form of digital rights management involving licenses to personal data, but then applied to enforce privacy. A similar approach is the use of \pdp{sticky policies}. Those could also be used to implement a right-to-be-forgotten (or at least a right-to-erasure).

\subsubsection{Strategy \#8: \pds{demonstrate}}

The final strategy, \pds{demonstrate}, requires a data controller to 
\begin{quote}
Be able to demonstrate compliance with the privacy policy and any applicable legal requirements.
\end{quote}
This strategy goes one step further than the \pds{enforce} strategy in that it requires the data controller to prove that it is in control. This is explicitly required in the new draft EU privacy regulation~\cite{COM(2012)11}.
In particular this requires the data controller to be able to show how the privacy policy is effectively implemented within the IT system. In case of complaints or problems, she should immediately be able to determine the extent of any possible privacy breaches, for example.

\paragraph{Design patterns}

Design patterns that implement this strategy are, for example, \pdp{privacy management systems}~\cite{DBLP:conf/trustbus/MontP05}, and the use of logging and auditing.

\section{Analysis}
\label{sec-validate}

\begin{figure}[t]
\begin{center}
\includegraphics[scale=\figscale]{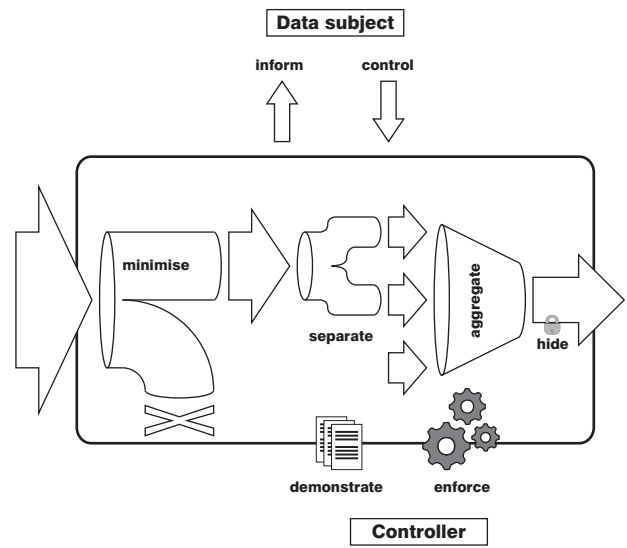}
\end{center}
\caption{The process flow metaphor of the eight privacy design strategies.} 
\label{fig-flow-metaphor}
\end{figure}

\subsection{Validation of our approach}

The eight privacy design strategies also naturally apply to an information flow system, as illustrated in Figure~\ref{fig-flow-metaphor}. In this view, \pds{minimise} corresponds to processing only a selected subset of the incoming data (and throwing the rest away), while \pds{separate} corresponds to splitting the data stream in 
several parts that are each further processed at separate locations. 
\pds{Aggregate} corresponds to combining (and compressing) data streams, while \pds{hide} (for example) encrypts the data while in transit. \pds{Inform}, \pds{control}, \pds{enforce} and \pds{demonstrate} are essentially the same as in the information storage model.

As a further validation of our approach, we verify that the eight privacy design strategies we derived cover the legal principles outlined in section~\ref{sec-legal}. Table~\ref{tab-mapping} shows this mapping. It was constructed by mapping the detailed description of each strategy in section~\ref{sec-pds} on each of the legal principles.

As discussed before, not every legal data protection principle can be covered by a privacy design strategy, simply because the design of the system has no impact on that principle. Some data protection principles, like purpose limitation, are only partially covered by some of the strategies. Realising purpose limitation in full also requires procedural and organisational means.

\newcommand{\rt}[1]{\rotatebox{90}{#1}}

\begin{table}[t]
\setlength{\tabcolsep}{5pt}
\begin{tabular}{l|cccccccccc}
 &
                  \rt{Purpose limitation} &
                      \rt{Data minimisation} &
                          \rt{Data quality} &
                              \rt{Transparency} &
                                  \rt{Data subject rights} &
                                      \rt{The right to be forgotten} &
                                          \rt{Adequate protection} &
                                              \rt{Data portability} &
                                                  \rt{Data breach notification} &
                                                      \rt{(Provable) Compliance}\\
\hline

\pds{minimise}    & o & + &   &   &   &   &   &   &   & \\ 
\pds{hide}        &   & + &   &   &   &   & o &   &   & \\ 
\pds{separate}    & o &   &   &   &   &   & o &   &   & \\ 
\pds{aggregate}   & o & + &   &   &   &   &   &   &   & \\ 
\pds{inform}      &   &   &   & + & + &   &   &   & + & \\ 
\pds{control}     &   &   & o &   & + &   &   & + &   & \\ 
\pds{enforce}     & + &   & + &   &   & + & + &   &   & o \\ 
\pds{demonstrate} &   &   &   &   &   &   &   &   &   & + \\ 
\end{tabular}\\
\vspace{1mm}\\
Legend:\\
+: covers principle to a large extent.\\
o: covers principle to some extent.
\caption{Mapping of strategies onto legal principles.}
\label{tab-mapping}
\end{table}

\subsection{Design pattern coverage}

With respect to design pattern coverage, we first observe that design patterns may belong to several design strategies. For example the \pdp{use pseudonyms} design pattern both implements the \pds{minimise} strategy and the \pds{hide} strategy. Similarly, a privacy enhancing technology may be applicable within several different privacy design patterns.

In the course of our investigations we also observed huge differences
between design strategies in terms of the number of design patterns (and corresponding privacy enhancing technologies) known to implement them. For the strategies \pds{minimise} and \pds{hide}, a large number of design patterns exist. This is not surprising, given the focus of most research in privacy enhancing technologies on these aspects of privacy protection. For the \pds{separate} and \pds{control} strategies on the other hand, no corresponding design patterns are known. This suggests further research on privacy design patterns should focus on these strategies.

\section{Conclusions}
\label{sec-concl}

We have defined the concept of a design strategy, and derived eight privacy design strategies based on data protection legislation, the OECD guidelines
and the ISO 29100 privacy principles. We have described these strategies in some detail, and have provided a first insight into the possible privacy design patterns that contribute to these privacy design strategies. Finally, we have validated our approach by verifying that the strategies cover the essential data protection principles.

We have taken the legal perspective as point of departure in our approach, and have validated our results against both the technological and the privacy policy perspective. We have not taken into account any philosophical, sociological or values-based perspectives. It would be interesting to investigate whether these perspectives have any impact on the list of privacy design strategies reported here.

This paper discusses work in progress. In particular, further research will be performed to classify existing privacy design patterns into privacy design strategies, and to describe these design patterns in more detail. Moreover, we have identified several implicitly defined design patterns (like \pdp{attribute based credentials}) that arise from our study of existing privacy enhancing technologies. 

Further developments and collaboration in this line of research will also be documented on our Wiki\footnote{%
  \url{http://wiki.science.ru.nl/privacy/}
}. 
We would very much welcome contributions from the research community.

\section{Acknowledgements}

I would like to thank the members of the Privacy \& Identity Lab\footnote{%
  \url{http://www.pilab.nl}
}
for discussions and valuable feedback. In particular I am grateful to Ronald Leenes, Martin Pekarek and Eleni Kosta for their detailed comments and recommendations that greatly improved this paper.

\bibliographystyle{plain}
\bibliography{pdp}

\end{document}